\documentclass[a4paper,11pt]{article} 
\usepackage{amsmath}
\usepackage{amssymb,amsfonts,textcomp}
\usepackage{graphicx}
\usepackage{hyperref}
\hypersetup{pdftex, colorlinks=true, linkcolor=blue, citecolor=blue, filecolor=blue, urlcolor=blue, pdftitle=, pdfauthor=, pdfsubject=, pdfkeywords=}

\newcommand{\comments}[1]{}

\title{Genetic Algorithm for determination of the event collision time and particle identification by time-of-flight at NICA SPD}
\date{$^1$Saint Petersburg State University, Laboratory of ultra-high energy physics, St.~Petersburg, Russia\\%
    $^2$Joint Institute for Nuclear Research, Dubna, Russia\\[2ex]}
\author{Semyon Yurchenko$^{1}$, Mikhail Zhabitsky$^{2}$\thanks{Correspondence: mikhail.zhabitskiy@cern.ch}\thanks{Submitted to MDPI Physics journal}
}

\begin{document}

\maketitle
\abstract{Particle identification is an important feature of the future SPD experiment at the NICA collider.
In particular, identification of particles with momenta up to a few $\text{GeV}/c$ by their time-of-flight will facilitate reconstruction of events of interest.
High time-resolution of modern TOF detectors dictates the need to obtain the event collision time~$t_0$ with comparable accuracy.
While determination of the collision time is feasible through the use of TOF signals supplemented by track reconstruction, it proves to be computationally expensive.
In this work we have developed a dedicated Genetic Algorithm as a fast and accurate method to determine the pp-collision time by the measurements of the TOF detector at the SPD experiment.
By using this reliable method for $t_0$ determination
we compare different approaches for the particle identification procedure based on TOF-signals.}



\section{Introduction}
The Spin Physics Detector (SPD) is a future experiment that will be placed in one of the two interaction points of the NICA collider in the Joint Institute for Nuclear Research. By studying collisions of polarized proton and deuteron beams,
the SPD collaboration will perform a comprehensive study of the unpolarized and polarized gluon content
of nucleons and other spin related phenomena~\cite{CDRSPD}.
With polarized proton-proton collision energies $\sqrt{s}$ up to $27~\text{GeV}$,
SPD will cover a kinematic range between the low-energy measurements at ANKE-COSY~\cite{Dymov:2016jgy} and SATURNE and the high-energy measurements at RHIC~\cite{STAR:2021mfd} and LHC~\cite{Hadjidakis:2018ifr}.

The SPD experimental setup is planned as a general-purpose $4\pi$~detector with advanced tracking and particle identification capabilities.
The particle identification will be performed
by means of $dE/dx$, Time-Of-Flight (TOF),
Electromagnetic calorimetry and Muon-filtering techniques.
The experiment will use a system of Multigap Resistive Plate Chambers (MRPC)~\cite{ZeballosMRPC,WangMRPC} as the TOF detector.
Its main aim is to provide $\pi/K/$p-identification of charged particles with momenta up to a few $\text{GeV}/c$. 

Identification of particles types by their time-of-flight is a well established technique in high energy physics collider experiments~\cite{ISRTOF, STARTOF, ALICETOF1, ALICETOF2, PANDATOF}.
It requires just three ingredients: $p$~-- momentum of the particle, $L$~-- arc length of its trajectory between the primary collision point and the TOF detector, $\tau$~--- the corresponding time-of-flight.
The latter is calculated as a difference between the stop and start ($t_{0}$) signals.
While the stop signal is measured with high precision by the TOF detector, the collision time $t_{0}$ cannot be obtained directly with the same accuracy.
The collision time can be estimated from timing of the accelerator or deduced from signals of a corresponding T0-detector, which will detect appearance of secondary particles scattered on small angles with respect to the beams axis,
but in this case uncertainty of the collision time will dominate uncertainty of the difference between stop and start signals.
Fortunately, the event collision time~$t_{0}$ can be determined with sufficient precision from the TOF measurements by means of $\chi^2$ minimization procedure~\cite{ISRTOF, ALICETOF2}.

As a significant number of secondary particles, originated from the proton-proton collision, subsequently enters the acceptance and is detected by the TOF detector, one can reconstruct $t_{0}$ as a common value for all detected particles through minimization of the sum of the squares of residuals.
A residual is defined as a difference between the measured TOF signal and its expected arrival time,
assuming a given mass hypothesis.
Thus the $\chi^2$ is minimised with respect not only to $t_0$ but to all mass hypotheses, which proved to be a difficult computational problem~\cite{ALICETOF1}.

As the bulk of secondary particles are pions, kaons or protons, it is natural to try different combinations of their masses in order to minimize $\chi^2$,
thus the minimization is performed over a discrete set of particles types.
The global minimum can be found by the Brute Force Algorithm (BFA), which is characterized by a very long run-time.
Authors have developed an Asynchronous  Differential Evolution-inspired~\cite{ADE_ACM} Genetic Algorithm (ADE-GA),
which solves the $\chi^{2}$ minimisation problem within significantly reduced computational time.

The reliable method for $t_{0}$ determination facilitates the identification of particles by their time-of-flight.
Several approaches can be used for the PID procedure~\cite{ALICEPID}.
In this work we compare performance of the Bayesian approach, "$n$-sigma" criteria and the direct solution of the $\chi^{2}$-minimisation problem.

\section{Time-Of-Flight detector and event selection}
\begin{figure}[ht]
    \centering
    \includegraphics[keepaspectratio, width=\textwidth]{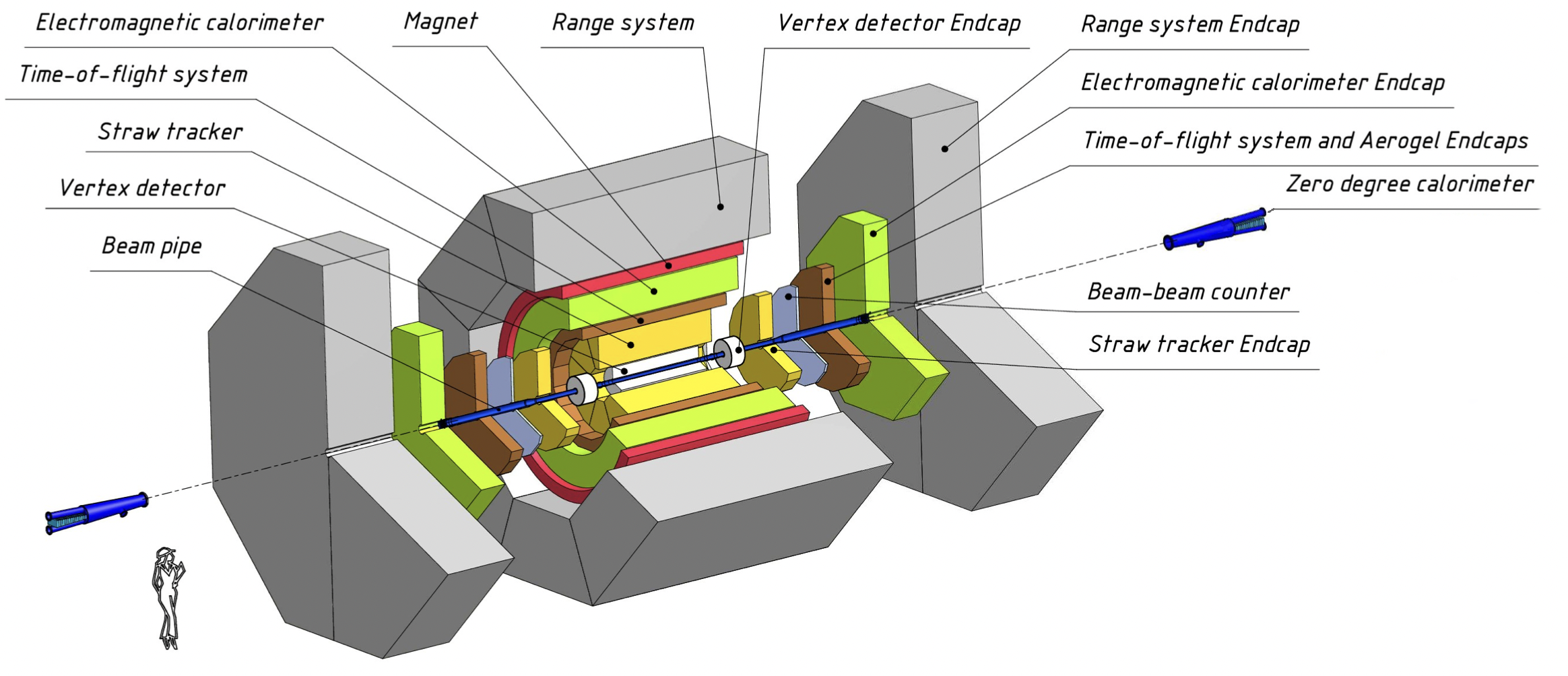}
    \caption{General layout of the SPD detector~\cite{CDRSPD}}
    \label{fig:spd_3d}
\end{figure}
The TOF system will consist of a barrel and two end-cap parts
with a radius of about 105~cm and a length of 370~cm.
It will have an overall active area of $27~\text{m}^2$ and cover polar angles greater than 100~mrad.
The short distance from the TOF detector to the interaction point dictates the TOF resolution to be within $50\div 60$~ps, which can be achieved with the MRPC technologies~\cite{WangMRPC}.
In our study we use a conservative estimation $\sigma_t=70$~ps.
The TOF detector will be located within the solenoidal magnetic field ($B=1$~T, parallel to the beams axis)
outside the inner tracker, which will measure momenta with relative precision
$\frac{\sigma_{p}}{p} = 2\%$~\cite{CDRSPD}.

PYTHIA8 Monte Carlo generator~\cite{PYTHIA83} has been used for simulation of proton-proton collisions at the center-of-mass energy $\sqrt{s}=27$\;GeV,
\textit{SoftQCD:all} settings have been selected to simulate minimum-bias events.
All charged particles have been propagated through the uniform magnetic field.
Intersection points of helix trajectories with the TOF detector have been calculated through the closed-form expressions.
Only charged tracks with momenta greater than $0.5~\text{GeV}/c$ have been used in the analysis, as relativistic particles are characterized by negligible track distortions due to multiple scattering and $dE/dx$-losses in the walls of the beam pipe and material of the inner tracker.
In the following the generated momentum value~$p_0$ and the arrival time $t_{i,0}$ to the TOF detector are smeared according to Normal distributions $p=N(p_0, \sigma_p)$ and $t_i=N(t_{i,0},\sigma_t)$. 

\section{Event collision time measurement performed by the TOF detector}
\subsection{Method to reconstruct the event collision time}
A large fraction of particles produced in proton-proton collisions at $\sqrt{s}=27~\text{GeV}$ has momenta below $2~\text{GeV}/c$,
which suggests a time-of-flight as a powerful technique for particles identification.
Statistically more than half of secondary charged particles detected by the TOF detector are pions.
Other major contributions are protons and charged kaons.
Admixtures of electrons and muons do not exceed a few percent and will be identified by the Electromagnetic Calorimeter and the Muon Range System~\cite{CDRSPD}.
One can calculate all possible times-of-flight
\begin{equation}\label{eq:tof_ij}
 \tau_{ij} = \frac{L_{ij}}{c} \sqrt{1+\frac{m_j^2}{p_i^2}},
\end{equation}
by assigning independently for each track~$i$ a certain particle type~$j$.
Formula~(\ref{eq:tof_ij}) works well only for relativistic particles,
for low momenta it has to be exchanged to piecewise summation along the particle's trajectory intersecting coordinate detectors.
Arc length $L_{ij}$ should take into account details of a type~$j$ particle interaction with matter. In this analysis we selected only tracks with momentum above $0.5~\text{GeV}/c$ where matter effects can be neglected.
For the event with $N$ reconstructed tracks the event collision time can be found by a $\chi^2$ minimization procedure:
\begin{equation}\label{eq:chi2}
    \chi^{2} = \sum_i^{N}\frac{(t_{0}+\tau_{ij}-t_{i})^{2}}{\sigma_{t}^{2}+\sigma_{\tau(j,p_{i})}^{2}} =
    \sum_i^{N}\frac{(t_{0}+\tau_{ij}-t_{i})^{2}}{\sigma_{ij}^2}.
\end{equation}
Where time-of-flight uncertainty $\sigma_{\tau(j, p_{i})}$ due to uncertainty in momentum is given by
\begin{equation}\label{eq:t0}
    \sigma_{\tau(j, p_{i})} = \frac{L_{ij}}{c}\cdot\frac{m^{2}_{j}}{p^{2}_{i}}\left(\sqrt{1+\frac{m^{2}_{j}}{p^{2}_{i}}}\right)^{-1}\cdot\frac{\sigma_{p}}{p}=0.02 \cdot\frac{L_{ij}}{c}\cdot\frac{m^{2}_{j}}{p^{2}_{i}}\left(\sqrt{1+\frac{m^{2}_{j}}{p^{2}_{i}}}\right)^{-1}.
\end{equation}
For pions with momenta $0.5~\text{GeV}/c$ it is as large as 100~ps and is taken into account.
Another contribution to $\sigma_{\tau(j, p_{i})}$ is uncertainty in the reconstructed track length.
In the SPD experiment tracking detectors will provide 30 to 40 hits per track in spatially separated detector planes~\cite{CDRSPD},
thus for $0.5~\text{GeV}/c$ pions time-of-flight uncertainty due to uncertainty in the reconstructed track length is less than 10~ps and is omitted in this work.

For a certain mass hypothesis an analytic solution for $t_{0}$ reads:
\begin{equation}\label{eq:chi2_t0}
t_{0}=\sigma_0^2\sum_i^{N}\frac{t_{i}-\tau_{ij}}{\sigma_{ij}^2},\quad\text{where}\quad \frac{1}{\sigma_0^2}= \sum_i^{N}\frac{1}{\sigma_{ij}^2}.
\end{equation}

So task is reduced to minimisation of $\chi^{2}$ (Eqs.~\ref{eq:chi2}--\ref{eq:chi2_t0}) by finding the proper mass hypothesis~--- vector of masses $(m_1,m_2,\ldots,m_N)$ for tracks in the event-by-event way.
The emphasis is paid to deduce an accurate and unbiased estimation of the collision time~$t_0$.
Sections \ref{subsec:brute} and \ref{subsec:gen} are dedicated for developing algorithms to perform the minimisation step.

\subsection{Brute Force Algorithm}\label{subsec:brute}
Most straightforward solution is to check all mass hypotheses and thus locate the global minimum~--- a combination of masses which has minimal $\chi^{2}$ (so-called exhaustive search or Brute Force Algorithm~--- BFA).
If $N_{m}$~--- number of possible masses (possible particles types), then the total number of combinations is $N_m^N$ and time complexity of this algorithm will be $O(N\cdot N_m^N)$.
Exponential running time means that this algorithm is computationally expensive if the number of reconstructed tracks exceeds 10.
To keep BFA execution time reasonable, possible particle types are restricted to $\pi^\pm$, $K^\pm$ and protons ($N_{m}=3$). 

\subsection{Genetic Algorithm}\label{subsec:gen}
The minimisation of $\chi^{2}$ (Eqs.~\ref{eq:chi2}--\ref{eq:chi2_t0}) is performed over discrete set of particle species, thus represents a typical problem in the domain of the discrete optimization.
To solve the problem we have developed an Asynchronous Differential Evolution-inspired~\cite{ADE_ACM} Genetic Algorithm (ADE-GA).

All possible types of particles are represented as an ordered by mass set (genetic representation),
e.g. $[m_{\pi},m_{K}, m_{p}]\rightarrow [0,1,2]$.
The algorithm maintains a set of candidate solutions called a population.
It optimizes a problem by iteratively improving the population through generation of new candidate solutions, which can replace inferior population members by means of natural (Darwinian) selection.
Opposite to the exhaustive search, the algorithm does not check all possible mass combinations, but identifies better solutions and performs further searches around them.  

Algorithm's workflow for an event with $N$ tracks is as follows:
\begin{enumerate}
    \item Create an initial population of $N_{pop}$ random candidates solutions.
    Each candidate solution is a random set of $N$ masses associated with corresponding tracks,
    each species has equal probability $\frac{1}{N_m}$ to be assigned to a given track.
    Initialization procedure enforces that all population members are unique and for each track there are at least two different masses within the population.\\
    Example of a population in event with 5 tracks and size of population $N_{pop}=5$:
    \vspace{6pt} \\
    \begin{tabular}{c|c}
        $v_{1}$ &  $(0,1,1,2,0)\leftrightarrow (m_{\pi},m_{K},m_{K},m_{p},m_{\pi})$\\
        $v_{2}$ &  $(2,2,0,1,0)\leftrightarrow (m_{p},m_{p},m_{\pi},m_{K},m_{\pi})$\\
        $v_{3}$ &  $(1,1,1,0,0)\leftrightarrow (m_{K},m_{K},m_{K},m_{\pi},m_{\pi})$\\
        $v_{4}$ &  $(0,0,0,2,1)\leftrightarrow (m_{\pi},m_{\pi},m_{\pi},m_{p},m_{K})$\\
        $v_{5}$ &  $(2,0,1,1,2)\leftrightarrow (m_{p},m_{\pi},m_{K},m_{K},m_{p})$\\
    \end{tabular}
    \vspace{6pt}
    \item Create a new candidate solution (offspring generation):
    \begin{enumerate}
        \item\label{step:offspring} Choose three distinct random solution vectors from the current population and create a mutant vector:
\begin{equation}{\label{eq:de_diff}}
    v_{mut} = v_{p} + (v_{r}-v_{q}).
\end{equation}
Vector $v_p$ is called a parent vector.
Two other vectors form a difference vector.
If any coordinate falls outside the range $[0, N_m-1]$, it is projected back to the corresponding boundary.
The mutant vector has to be different from any population member, otherwise the generation is repeated.\\
Example: $v_{p}=(0,1,1,2,0)$, $v_{r}=(2,2,0,1,0)$, $v_{q}=(1,1,1,0,0)$,
\begin{equation}
    v_{mut}=(0,1,1,2,0) + (2,2,0,1,0)-(1,1,1,0,0)=(1,2,0,2,0).
\end{equation}
\item Calculate fitness of the offspring: $t_{0}^{mut}$ and $\chi^{2}_{mut}$ (see Eqs.~\ref{eq:chi2}--\ref{eq:chi2_t0}),
\item Compare $\chi^{2}_{p}$ and $\chi^{2}_{mut}$,
\item\label{step:darwin} If $\chi^{2}_{mut}<\chi^{2}_{p}$~--- the new mutant vector is better than the parent, then the offspring supersedes the parent vector in the population.
Otherwise the population remains unchanged.
This step is called natural (Darwinian) selection.
    \end{enumerate}
\item 
Steps \ref{step:offspring}--\ref{step:darwin} are repeated until a stop criteria is met.
After a predefined number of iteration $N_{steps}$, the solution with the smallest $\chi^{2}$ is chosen as the best combination.
\end{enumerate}

This algorithm has only one control parameter~--- size of the population $N_{pop}$.
Numerical simulations proved that $N_{pop}=15$ is sufficient to solve the problem.
Time complexity of Genetic Algorithm is $O(N\cdot N_{pop}\cdot N_{steps})$, where $800<N_{steps}<1000$.
Time complexity increases only linearly as a function of track number, which makes the algorithm suitable for high-multiplicity events.
Also the Genetic algorithm is not limited to $N_m=3$, but can perform the global search for a wider range of possible particle types without loss of performance. 

Besides the population size $N_{pop}$ the canonical Differential Evolution has two other control
parameters: crossover rate $C_r$ and scale factor $F$.
In this study $C_r=1$ because minimization variables are correlated, $F=1$ is chosen due to granularity of the mass spectra. 
The developed ADE-GA algorithm represents the asynchronous Evolutionary Algorithm: it updates randomly selected population members
by the ADE/rand/rand/1 strategy~\cite{ADE_ACM}.
In this approach fitness of many candidate solutions can be calculated in parallel, which will further speed-up calculations.

The algorithm can be further accelerated if one monitors the convergence speed and carefully chooses termination (stop) criteria.
Results, cited in this work, were obtained with the algorithm performing a predefined fixed number of iterations~$N_{steps}$.
In this approach the maximal number of allowed iterations~$N_{steps}$ is chosen to \emph{guarantee} a high convergence rate to the global minimum.
Analysis of the convergence shows that for most events the minimum is found by a much earlier iteration and further iterations waste computing time.
In the following, several approaches towards early detection of global convergence are discussed.

Result of successive iterations of the ADE-GA algorithm is a gradual improvement of the population: naturally selected candidate solutions have smaller $\chi^2$-values than their respective parent vectors.
Not only the best vector, but all population members converge to the minimum.
Thus the spread in fitness function values within the population is gradually reduced, the small spread can indicate either convergence or stagnation of the algorithm~\cite{ADESTOP}.
To monitor convergence one can sort all population members by their fitness values: $\chi^2_{best},\ldots,\chi^2_{m},\ldots,\chi^2_{worst}$, where $\chi^2_{m}$ denotes the median fitness.
As the fitness of the global minimum is expected to be of the order of $N$, where $N$ is a number of tracks, one can stop iterations as soon as $(\chi^2_{m} - \chi^2_{best})/N<\Delta_m$, where $\Delta_m$ is a predefined small value.
Alternatively, algorithm can monitor typical number of iterations between successive improvements of the $(\chi^2_{m} - \chi^2_{best})$ difference~--- $N_{progress}$, which can be achieved through learning in the process.
If there is no progress after $kN_{progress}$ iterations, where typically $k=3\ldots 5$, then the algorithm is terminated.
Monitoring the $(\chi^2_{m} - \chi^2_{best})$ difference has major advantages with respect to stop criteria based only upon the $\chi^2_{best}$-value.
The improvement steps by the algorithm can be characterized by their exploration or exploitation chances.
Exploration is the ability of the algorithm to locate a new region  in the search domain with better fitness values.
Exploitation is a gradual improvement of the population through testing of potentially interesting candidate vectors around an already found local minimum.
Differential Evolution is well known for its exploration abilities.
As soon as a new region of interest is located, the algorithm quickly populates the neighborhood of the local minimum, thanks to the mutation operator (Eq.~\ref{eq:de_diff}).
Improvement steps by exploration are a much rare case, while improvements through exploitation are common.
If one monitors only the $\chi^2_{best}$-value as a stop criterion, then after a successful exploration step one can cause a premature termination of iterations by preventing a further fast exploitation around a new minimum.
Typically at this stage exploitation leads not only to a general improvement of the population but also to finding a better best-so-far solution.

\section{Results and Discussion}

\subsection{Comparison of the Genetic Algorithm with the Brute Force Algorithm}\label{subsec:comp}
\begin{figure}
    \centering
    \includegraphics[keepaspectratio, width=\textwidth]{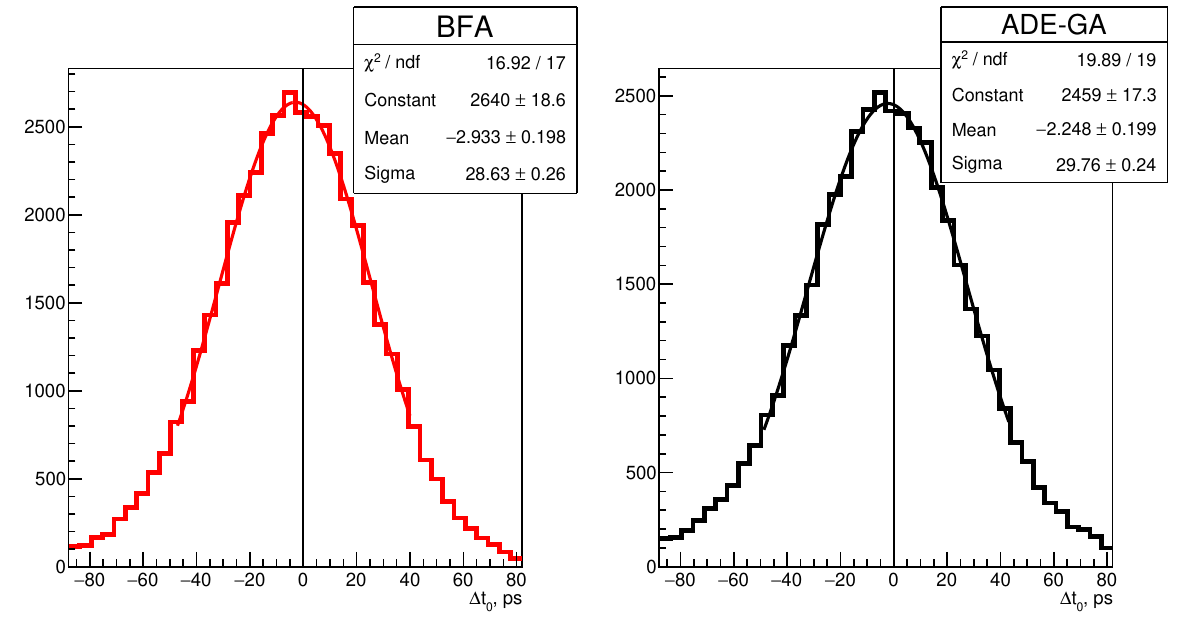}
    \caption{Distributions of errors $\Delta t_0 = t_{0}-t^{true}_{0}$ for the event collision time reconstructed by the Brute-Force algorithm (left) and by the DE-inspired Genetic Algorithm (right)}
    \label{fig:tzeros}
\end{figure}

Brute Force Algorithm finds the global minimum of $\chi^{2}$ minimisation and is used as a reference to check performance of the Genetic Algorithm.
Due to high time complexity we can use Brute Force Algorithm as a reference only in events with low multiplicities ($5\leq N\leq14$).
Distributions of errors $\Delta t_0 = t_{0}-t^{true}_{0}$ for such events are presented in Fig.~\ref{fig:tzeros}.
Only $\pi^\pm, K^\pm, p^\pm$ are used as allowed particle types.
Both algorithms provide unbiased estimation of the reconstructed event collision time with resolutions of $29$\;ps for the Brute Force Algorithm and $30$\;ps for the Genetic Algorithm.

Another important metric is the overall percentage of tracks that were identified correctly: $97.2\%$ for the Brute Force Algorithm, $96.8\%$ for the Genetic Algorithm.
Non-zero PID inefficiency by BFA looks counter-intuitive, but it appears due to finite resolution of the TOF detector when uncertainty of its  measurement exceeds typical time-of-flight difference between two different particle types at a given momentum.
In this case a particle will be misidentified if the global minimum of $\chi^2$ minimization is deeper than the $\chi^2$ of the actual particle configuration.

\begin{figure}
    \centering
    \includegraphics[keepaspectratio, width=\textwidth]{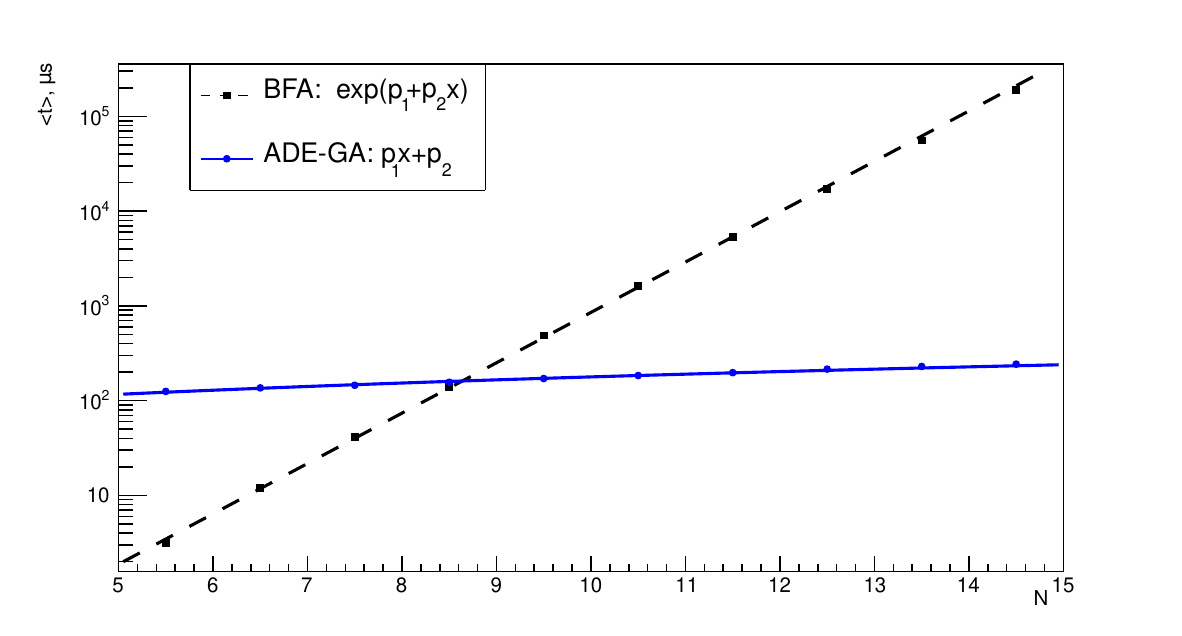}
    \caption{Time complexity comparison of the Brute-Force and the DE-inspired Genetic algorithms:
    the median run-time $\langle t\rangle$ as a function of event multiplicity}
    \label{fig:time}
\end{figure}

The Genetic Algorithm performance to solve $\chi^2$ minimization problem is on a par with the exhaustive search,
but it demonstrates a different time complexity for high-multiplicity events (see Fig.~\ref{fig:time}).
While for events with less than 8~tracks Brute Force Algorithm has shorter run time, it exponentially slows down as multiplicity grows.
Average run time of Brute Force Algorithm on events with $5\leq N \leq14$ is $5$\;ms, while Genetic Algorithm is much faster~--- $160\;\mu$s.
Both BFA and ADE-GA are intrinsically parallel algorithms.
Run-times, cited in this article, have been measured in a single-thread calculation mode to simplify the comparison.
Faster execution time is achieved with multithreading. 

Analysis by ADE-GA of events with any number of reconstructed tracks confirms that the uncertainty in the collision time $\sigma_{0}$
decreases from about $32$\;ps for 5-track events down to about $20$\;ps for high-multiplicity events, it scales as $1/\sqrt{N}$.
The achieved uncertainty in $t_0$ is much better than the resolution of the TOF detector $\sigma_t=70$\;ps, thus the latter will dominate uncertainty in the time-of-flight between the collision point and the TOF detector.
The efficiency of the ADE-GA to accurately reconstruct the collision time is estimated to be about 97\%.

Ability of the ADE-GA algorithm to efficiently solve the global minimization problem (Eq.~\ref{eq:chi2}) defined in a discrete space is based on the following fundamental principles of Differential Evolution~\cite{DE,DEReview}.
First, the algorithm is derivative-free, thus it can perform optimization over discrete variables.
Second, Differential Evolution doesn't use any assumption about a particular shape of the minimized fitness function, e.g. relying on the linear or quadratic approximation of its shape.
Instead, Differential Evolution adapts its population to a particular landscape through natural (Darwinian) selection.
Better candidate solutions have higher chances to stay in a population for a longer time thus more often playing the parent (central) role in the mutation operator ($v_p$ in Eq.~\ref{eq:de_diff}).
In this way the population center of gravity is gradually shifted to the deeper minimum in case of multimodal problems. 
Last but not least, due to common convergence of population members to a minimum, the algorithm automatically adapts the difference vector ($v_r-v_q$ in Eq.~\ref{eq:de_diff}) to a typical size of the search region around the minimum.
The latter feature enables the comparatively fast convergence of Differential Evolution. 

\begin{figure}
    \centering
    \includegraphics[keepaspectratio, width=0.5\textwidth]{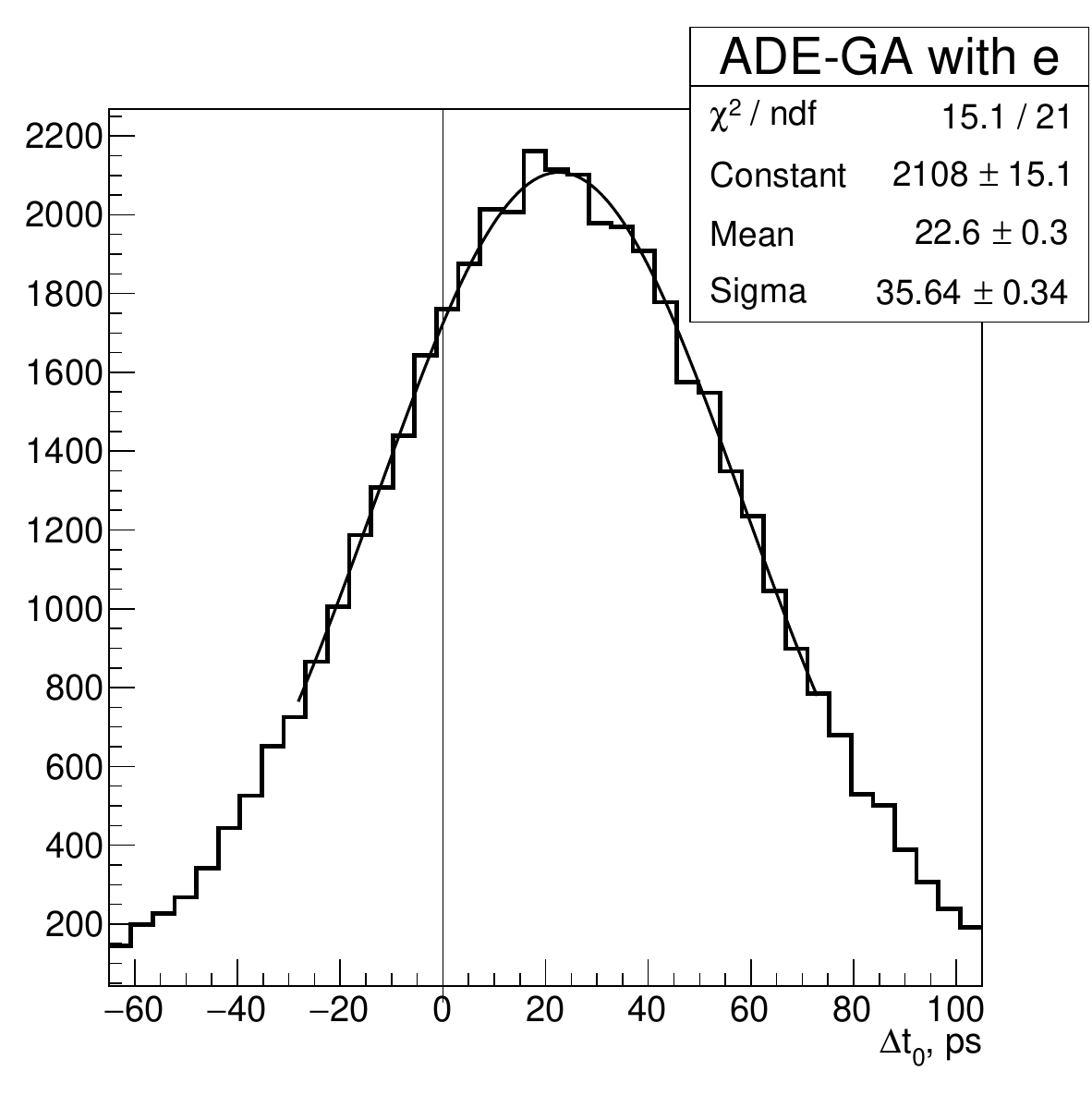}
    \caption{Distributions of errors $\Delta t_0 = t_{0}-t^{true}_{0}$ for the event collision time reconstructed by ADE-GA with $\mathrm{e}^\pm, \pi^\pm, K^\pm, p^\pm$ species}
    \label{fig:diffmass}
\end{figure}

Thanks to the swiftness of the Genetic algorithm, a wider than $N_m=3$ range of possible mass types can be taken into account. 
If electrons/positrons are added then the reconstructed event collision time becomes biased (Fig.~\ref{fig:diffmass}).
Due to short flight paths, the expected arrival time of pions with momenta above $1~\text{GeV}/c$ to the TOF detector is delayed to electrons less than the TOF time-resolution. 
In this case some pions will be misidentified as electrons whenever such a mass hypothesis provides a deeper minimum for $\chi^{2}$.
As pions are much more abundant than electrons/positrons such misidentification will result in a biased estimation of the collision time.

To remedy the probability of misidentifications one can remove from consideration any track which type is not identified in a unique way for sure (for example by $n$-sigma criteria, see below).
This approach was derived in work~\cite{POLINA}: using a priori knowledge about dominant prevalence of pions in the sample of registered particles, one can consider all low momenta tracks as pions, for each track calculate estimation of the collision time, identify the most probable one and reject all heavier-than-pion particles.
In this way the collision time can be found with uncertainty about $32\;$ps for events with a fairly high number of tracks, but not in events with less than 3~pions below $1.5~\text{GeV}/c$.

As the accurate and unbiased estimation of the event collision time is the main goal of this study, the correctness of the obtained $t_0$ value is further verified by iteratively removing major addends from the $\chi^2$-sum (Eq.~\ref{eq:chi2}) followed by the $\chi^2$ minimization over the rest of tracks in the event.
Statistically significant shift of the $t_0$ value indicates a possible outlier due to noise or misidentification. 

\subsection{Alternative ways to measure the event collision time}
Alternatively the event collision time can be measured by dedicated detectors installed close to the beam tube, so-called T0 detectors~\cite{ALICE_T0}.
Such detectors typically have fine granularity to cope with a high load of secondary particles and protons scattered on small angles.

In SPD the intersection region of two colliding beams will cover a few tens centimeters along the beam axis.
This dictates the necessity to install a pair of T0 detectors, located from both beam directions around a collision point, to be used in combination.
The Monte-Carlo simulation shows: if T0 detectors cover polar angles between 60 and 500~mrad then only about half of proton-proton collisions at $\sqrt{s}=27~\text{GeV}$ will produce charged tracks in both forward and backward T0 detectors~\cite{CDRSPD}.
This limits the ability of T0~detectors to measure the event collision time in the SPD experimental conditions.
At the same time T0~detectors can determine $t_0$ for events in which other detectors can not be used, e.g. in the case of elastic scattering.

In case of T0~detectors only few tracks are involved in the event collision time determination, while $t_0$~measurement by the TOF detector uses many tracks and improves as $1/\sqrt{N}$ for high-multiplicity events.
Moreover measurements by the TOF detector are accompanied by reconstructed tracks which further reduce uncertainties due to uncertainties in time-of-flight distance and particle momentum.

\subsection{Particle identification by time-of-flight}
As the reliable method to reconstruct the event collision time $t_{0}$ is developed one can perform particle identification through comparison of track timing by the TOF detector to the expected time of particle's arrival to the detector. 
There are several strategies for PID by time-of-flight:
\begin{enumerate}
    \item One can assign particle type for each track from the result of $\chi^{2}$-minimization:
    the track type is accepted as the most likely species (maximal probability).
    \item Or for every track~$i$ in event one can exclude it from determination of the collision time~$t_{0}$ to avoid correlations.
    Let's denote as $t_{i0}$ the event collision time calculated over the rest of tracks in the event.
    Then there are two common strategies to perform PID by time-of-flight~\cite{ALICEPID}:
    \begin{enumerate}
        \item $n$-sigma selection~--- the most simple threshold discriminator: \begin{equation}
            n_{ij} = \frac{t_i - (t_{i0} + \tau_{ij})}{\sigma_{ij}}
            = \frac{S_{i}-\hat{S_{i}}(m_{j})}{\sigma_{ij}}.
        \end{equation}
        Here $S_{i}$ is a signal obtained for track~$i$, $\hat{S}_{i}(m_{j})$ is the expected signal for a particle of species~$j$ with momenta~$p_i$.
        If the signal belongs to the range $\pm2\sigma$ or $\pm3\sigma$ of a certain species this track is accepted as the particle of this species.
        Track can be accepted as multiple species. 
        \item Bayesian method: takes into account yield of particle species.
        The conditional probability for track~$i$ to be a particle of species~$j$ reads:
        \begin{equation}\label{eq:bayes}
            P(H_{j}|S_i)=\frac{P(S_i|H_{j})C(H_{j})}{\sum_{\alpha = \pi,K,p}P(S_i|H_{\alpha})C(H_{\alpha})}.
        \end{equation}
        Here $C(H_{j})$ is a prior probability that is calculated iteratively.
        It takes into account relative abundance of species~$j$, which depends on particle momenta and emission angle.
        The likelihood function $P(S_i|H_{j})$ is given by:
        \begin{equation}
            P(S_i|H_{j}) = \frac{1}{\sqrt{2\pi}\sigma_{ij}}\exp \left(-\frac{1}{2}n^{2}_{ij}\right).
        \end{equation}
   \end{enumerate}
\end{enumerate}

\begin{figure}
    \centering
    \includegraphics[width=\textwidth]{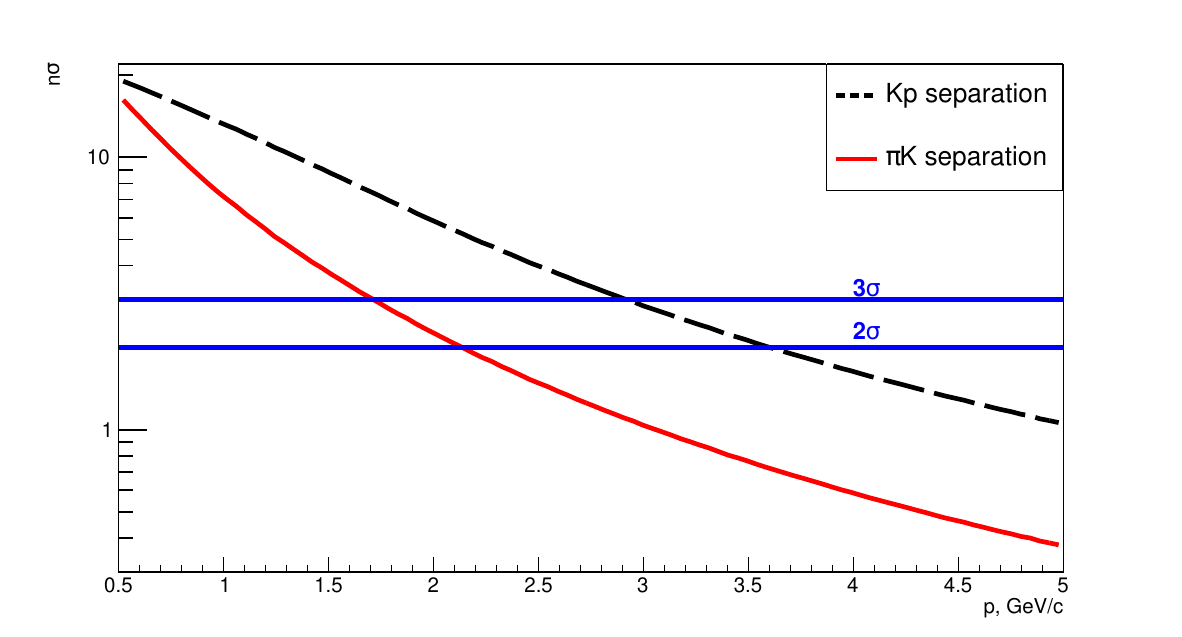}
    \caption{$\pi/K$ and $K/$p separation powers as functions of momenta}
    \label{fig:seppower}
\end{figure}

Separation power $n\sigma_{\pi K}=(\tau_{iK} - \tau_{i\pi})/\sigma_{iK}$ can be used as a measure of the PID performance~\cite{ALICETOF2}. 
In the SPD, identification of particles by their time-of-flight can be performed up to $1.7~\text{GeV}/c$ for $\pi/K$ separation and up to $3~\text{GeV}/c$ for $K/$p
at $3\sigma$ level (Fig.~\ref{fig:seppower}).

\subsection{PID benchmarks for two-prong decay channels}

Different PID methods have been compared by reconstructing several decay channels with two oppositely charged particles in the final state:
$\phi\rightarrow K^{+}K^{-}$, $\Lambda\rightarrow p^{+}\pi^{-}$ and $K^0_{s}\rightarrow \pi^{+}\pi^{-}$.
For this study two-prong decay channels have been chosen due to smaller combinatorial background with respect to multi-prong decays.
Only PID by time-of-flight is used in this section.
One should note that in real data analysis it will be accompanied by other methods to reduce background: secondary vertex reconstruction of intermediate particles, particle identification by $dE/dx$ and so on.

\begin{figure}
    \centering
    \noindent%
        \includegraphics[keepaspectratio, width=0.49\textwidth]{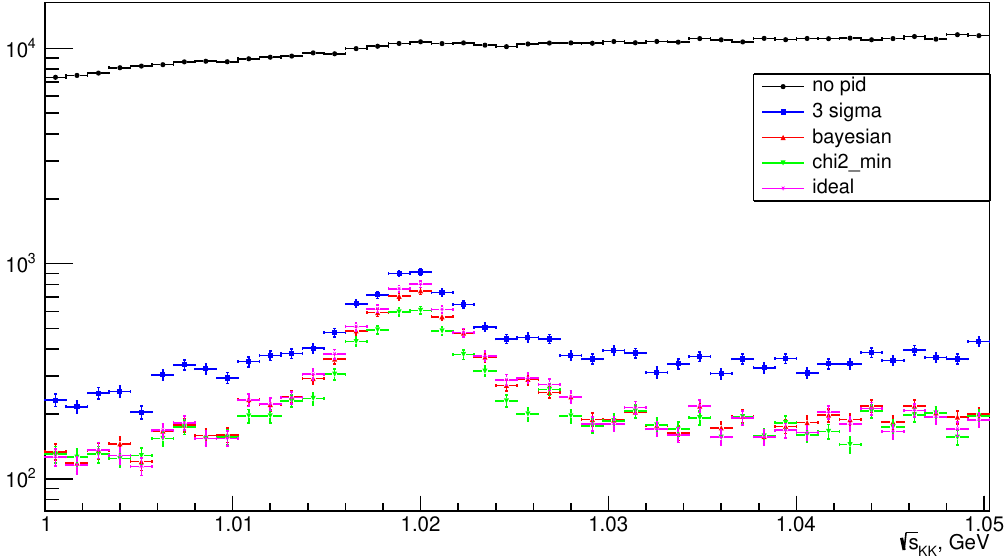} 
        \hfill%
        \includegraphics[keepaspectratio, width=0.49\textwidth]{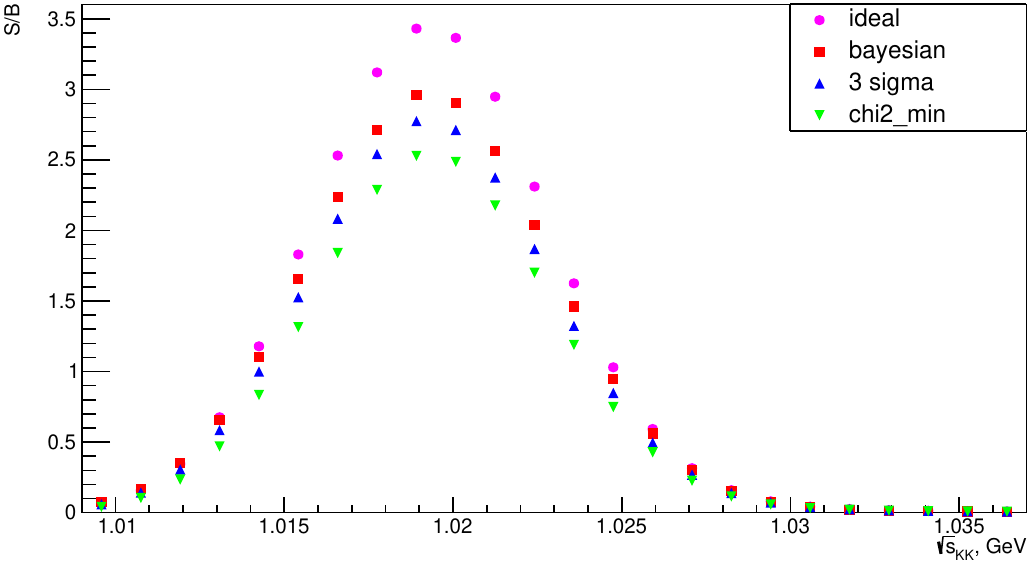}
    \caption{Invariant mass of pairs of oppositely charged tracks assumed to be $K^+K^-$ (left) and corresponding
    signal-to-background ratios (right) with different PID strategies}
    \label{fig:phimass}
\end{figure}

\begin{figure}
    \centering
    \noindent%
        \includegraphics[keepaspectratio, width=0.49\textwidth]{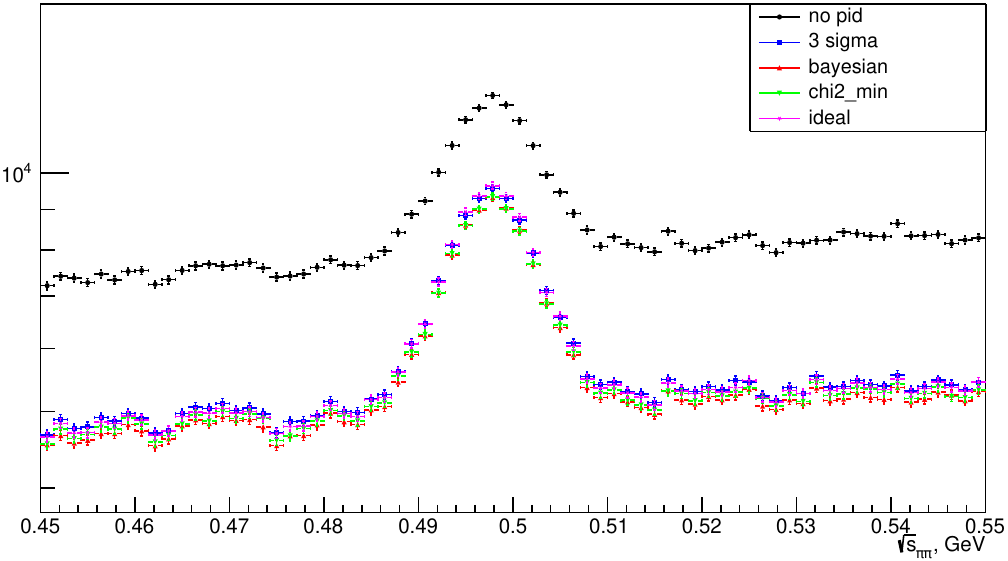} 
        \hfill%
        \includegraphics[keepaspectratio, width=0.49\textwidth]{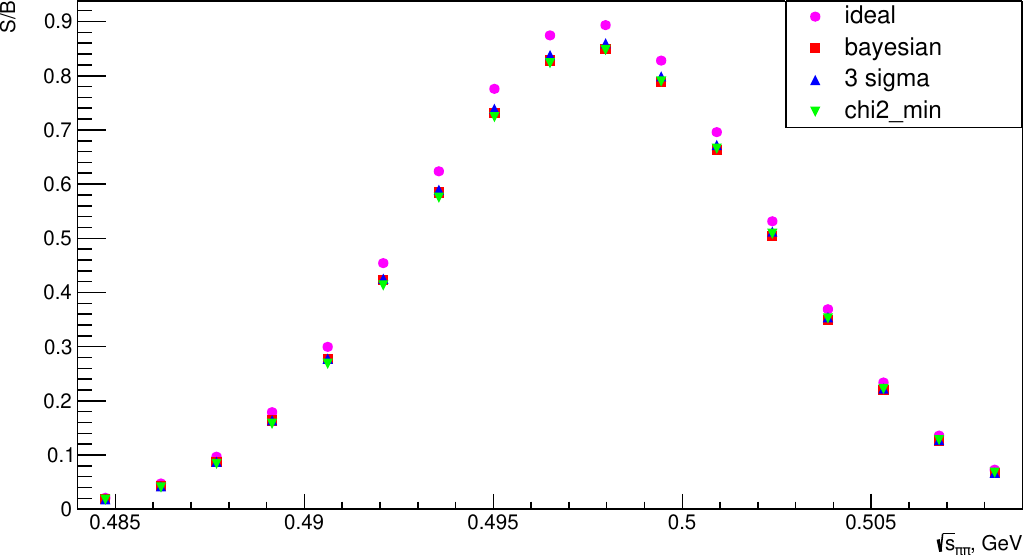}
    \caption{Invariant mass of pairs of oppositely charged tracks assumed to be $\pi^+\pi^-$ (left) and corresponding
    signal-to-background ratios (right) with different PID strategies}
    \label{fig:kshort}
\end{figure}

\begin{figure}
    \centering
    \noindent%
        \includegraphics[keepaspectratio, width=0.49\textwidth]{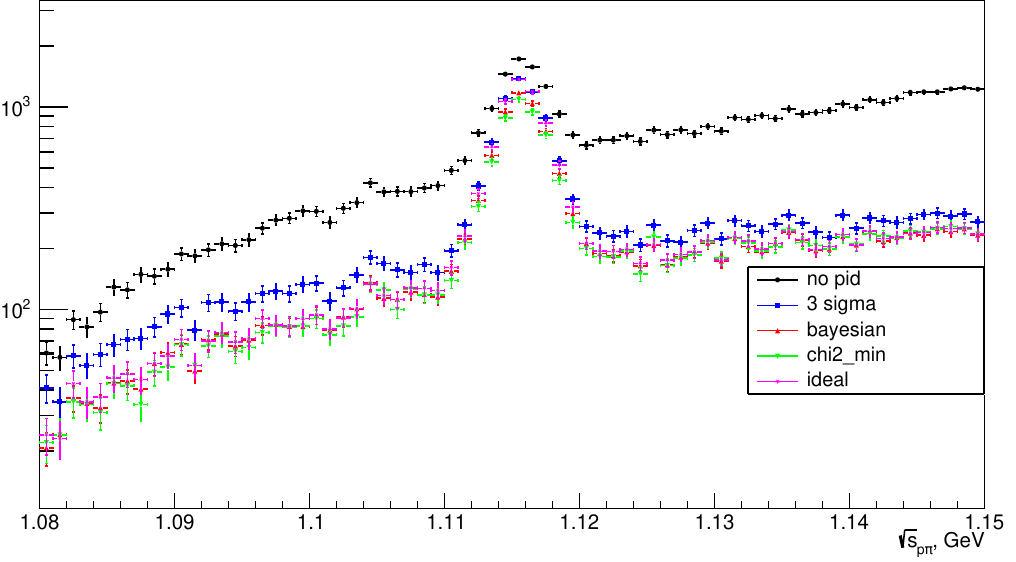} 
        \hfill%
        \includegraphics[keepaspectratio, width=0.49\textwidth]{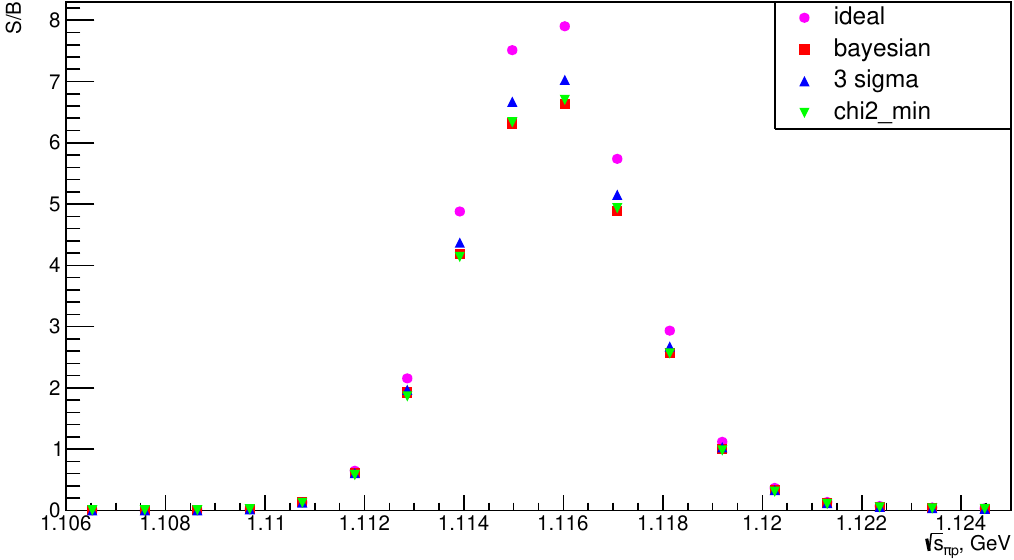}
    
    \caption{Invariant mass of pairs of oppositely charged tracks assumed to be p$\pi^-$  (left) and corresponding
    signal-to-background ratios (right) with different PID strategies}
    \label{fig:lambda}
\end{figure}

In Figs.~\ref{fig:phimass}--\ref{fig:lambda} invariant mass of all pairs of oppositely charged tracks is shown.
The same combination of masses has been assigned to each pair as for the channel of interest ("no pid" in figures).
In the $n$-sigma approach only tracks which have TOF signals within $\pm3\sigma_{ij}$ of a certain species~$j$ are selected ("$3$~sigma").
In case of the weighted Bayesian PID all combinations are included with the pair's weight as a product of conditional probabilities~(Eq.~\ref{eq:bayes}) for each prong ("bayesian").
The possible pair combinations corresponding to the global minimum of the $\chi^2$-minimization are marked as "chi2\_min".
Finally, "ideal"\ corresponds to Monte-Carlo combinations with known particle types.

The weighted Bayesian approach, which exploits both PID-by-TOF capabilities and abundance of particle species,
provides the best suppression of background while preserving particles of interest.
Kaons are less abundant then pions and protons,
therefore the advantage of the Bayesian approach is more pronounced if kaons are the decay products (Fig.~\ref{fig:phimass}).
By applying the $n$-sigma approach one can preserve more signal events, but at the same time the suppression of the combinatorial background is reduced. 
Benchmarks analyses, shown in Figs.~\ref{fig:phimass}--\ref{fig:lambda},
demonstrate the power of the identification of particles by the time-of-flight method in the SPD experimental conditions.

Study of open-charm $D$-mesons production is one of the main goals of the SPD experiment.
$D$-mesons preferentially decay by weak interaction into kaons and pions.
In the case of $D^{0}\rightarrow \pi^{+} K^{-}$ (and charge conjugate) decays, the Bayesian PID by time-of-flight allows to suppress combinatorial two-prong background by a factor~15,
thus enhancing the analysis of open-charm decays.

\section{Conclusions}
In the SPD experiment the accurate determination of the event collision time is required to perform $\pi/K/$p identification in low momenta range ($0.5{-}3~\text{GeV}/c$) by the time-of-flight method.
The collision time can be reconstructed on an event-by-event basis by a minimization procedure, which uses measurements of the TOF detector combined with track reconstruction,
but the corresponding solution is computationally expensive.
In this work we present a dedicated Asynchronous Differential Evolution-inspired Genetic Algorithm which solves the optimization problem in the direct way without any simplifications, thus providing the fast and reliable measurement of the event collision time throughout the range of track multiplicities.
Finally, different strategies of particle identification by time-of-flight are tested to prove the power of the PID-by-TOF method in the SPD experimental conditions.

\vspace{6pt} 

\section*{Author contributions}
Conceptualization, supervision: Mikhail Zhabitsky; formal analysis, methodology, software, investigations: Semyon Yurchenko. Both authors have read and agreed to the published version of the manuscript.
\section*{Funding}
This work was in part supported by the JINR START 2022 program.
\section*{Acknowledgments}
The authors would like to express their deepest appreciation to members of the SPD Collaboration for their valuable feedback.
\section*{Conflict of interest}
The authors declare that the research was conducted in the absence of any commercial or financial relationships that could be construed as a potential conflict of interest.

\section*{Abbreviations}
The following abbreviations are used in this manuscript:\\

\noindent 
\begin{tabular}{@{}ll}
ADE-GA & Asynchronous Differential Evolution-inspired Genetic Algorithm\\
BFA & Brute Force Algorithm\\
DE & Differential Evolution\\
GA & Genetic Algorithm\\
MRPC & Multigap Resistive Plate Chamber\\
NICA & Nuclotron-based Ion Collider fAcility\\
PID & Particle IDentification\\
SPD & Spin Physics Detector\\
TOF & Time-Of-Flight\\
\end{tabular}


\end{document}